# Reproducibility in Computational Materials Science: Lessons from 'A General-Purpose Machine Learning Framework for Predicting Properties of Inorganic Materials'

Daniel Persaud[a,b], Logan Ward[c], Jason Hattrick-Simpers[a,b*]

[a]Department of Materials Science and Engineering, University of Toronto, Canada
[b]Acceleration Consortium, University of Toronto, Canada
[c]Data Science and Learning Division, Argonne National Laboratory, USA

The integration of machine learning techniques in materials discovery has become prominent in materials science research and has been accompanied by an increasing trend towards open-source data and tools to propel the field. Despite the increasing usefulness and capabilities of these tools, developers neglecting to follow reproducible practices creates a significant barrier for researchers looking to use or build upon their work. In this study, we investigate the challenges encountered while attempting to reproduce a section of the results presented in "*A general-purpose machine learning framework for predicting properties of inorganic materials*." Our analysis identifies four major categories of challenges: (1) reporting computational dependencies, (2) recording and sharing version logs, (3) sequential code organization, and (4) clarifying code references within the manuscript. The result is a proposed set of tangible action items for those aiming to make code accessible to, and useful for the community.

* Correspondence: jason.hattrick.simpers@utoronto.ca

## 1. Introduction

Ensuring reproducibility when applying machine learning (ML) is an important, yet overlooked aspect of research, leading some to consider it in a state of "crisis" [1]–[3]. Computational reproducibility is to obtain consistent results with the original work using the same methods, code, and data [4]. These three components play a fundamental role in material informatics (MI), a field where statistical methods, ML, and materials data are used to understand processing-structure-property-performance (PSPP) relationships [5]. MI represents a paradigm shift in materials science research, shifting from the expert-driven empirical approach, towards data-driven techniques to understand the PSPP relationship [6]–[13]. As MI continues to grow, so too does the development of novel tools and platforms to support implementation [14]–[17]. However, a gap persists in developers adopting reproducible practices when distributing MI tools, hindering the efforts of researchers looking to implement them. Despite concerns about a developer's embedded bias, known or unknown, propagating to users [18], promoting reproducibility of tools instills trust and facilitates easy adoption.

An early seminal contribution to the field of MI was presented by Ward et al. in 2016 [14], introducing a unique ML framework to predict properties of inorganic materials. Despite the significance of the framework, which has subsequently been extended and further developed, we found the version of the code provided with in the supplementary information (SI) to be incomplete. Specifically, the SI only included a text interface script to build the proposed framework, referred in this study as the *model building* script, while any scripts for extending the framework, referred in this study as the *extensibility analysis*, were entirely missing. Since some scripts are unavailable, the objective of this study is to replicate (obtain consistent results using new data or methods [4]) the published results. We will provide an account of the methodology from the original work, describe the difficulties encountered during the replication, and explain how we address these challenges. Building on this, we propose a set of recommendations to enhance the reproducibility of future open-source tool development and, thereby, promote reliability and transparency in published MI results.

## 2. Summary of Ward et al. 2016

The intention of Ward et al. was to create a framework for building ML models which could be applied to a wide variety of inorganic material problems. The central outcome was the Material Agnostic Platform for Informatics and Exploration (Magpie) descriptors, a method of obtaining a broad range of chemically meaningful quantitative descriptors for an inorganic compound based on its elemental composition. In addition to the Magpie descriptors, their proposed framework incorporated a hierarchical modeling approach to group data into chemically similar sections, then train a model on each subset, reducing the breadth of the descriptors per model. Part of the paper focused on demonstrating the effectiveness of Magpie in solar cell design, and we seek to recreate it from only the description, software, and data provided with the original publication.

The SI that was provided with the Ward et al. paper contained the source code and compiled version of the associated Magpie software; which is written in Java [19], a set of input scripts related to the main parts of the paper, data needed to train the ML models, and documentation from the author. The scripts describe how to use Magpie to solve a specific problem (e.g., build a machine learning model with specific parameters) and are written in a text interface language which is interpreted by Magpie. In theory, these are sufficient to recreate the paper in its entirety because they contain the same ingredients (software, data, inputs) used by the original authors. However, the SI lacks the exact scripts used for the *extensibility analysis* presented in the paper.

A first key claim in Ward et al. involved predicting bandgap energies of crystalline compounds from an ensemble of Reduced Prunning Error Trees (REPTrees) then comparing it to a single REPTree and random selection. The novel hierarchical strategy made 67% accurate predictions, outperforming the single REPTree and random selection, which yielded only ~46% and ~12% respectively. The performance difference between the hierarchical REPTrees and the single REPTree legitimized the concept, and the *model building* script was made available in the SI.

The framework was then extended to address a practical material design problem: identifying new materials with bandgaps suitable for solar cells from a set of unexplored ternary compounds. The outcome of this task was a list of five potential candidate compounds, along with their predicted bandgap (table 1). The model to make these predictions are available from the *model building* script, but instructions on how to train it and then use it to make new predictions were not. Given the unavailability of the original code and the lack of validation for these predictions, we focused on replicating this *extensibility analysis*.

Table 1: Compositions and their predicted bandgap energies from the original work [14].

| Composition | $E_g$ (eV) |
|---|---|
| $ScHg_4Cl_7$ | 1.26 |
| $V_2Hg_3Cl_7$ | 1.16 |
| $Mn_6CCl_8$ | 1.28 |
| $Hf_4S_{11}Cl_2$ | 1.11 |
| $VCu_5Cl_9$ | 1.19 |

## 3. Replication of *Search for New Solar Cell Materials* and Gaps in Reproducibility

Ward et al. made significant efforts to promote reproducibility by providing raw data, a script to build the proposed hierarchical model, as well as extensive documentation, but even still it is challenging. We detail the challenges encountered at each stage of replicated the results.

### 3.1. Installation and Recreating Scripts

Our first problem was installing the dependencies needed to run Magpie. The requirements are not just buried several pages deep in documentation but are also misleading. The documentation states that Magpie requires "Java Runtime Environment (JRE) Version 7 (v7) or greater", but it is incompatible with the current Java version (v20). Moreover, JREv7 is no longer supported by Oracle, lacks security patches, and is only available for developers. While this is a slight challenge, it is overcome by installing the still-supported Java v8, but the misleading requirements hinder reproducibility.

As noted previously, we also had to resort to recreating *extensibility analysis* scripts that were not made available on publication. Learning the input language for Magpie presents a first barrier to reproducing the results from Ward et al. Building these scripts also requires relying on imprecise, human language descriptions from the manuscript. There are many points for deviation between original study and software, for example ambiguous references to ML models ("our model" vs. "the hierarchical model"), and data (OQMD vs the ICSD entries in OQMD) without clarity on the specific entities being referenced. Disparity between documented methodologies and their practical implementation in code underscores a notable gap. We recreated the scripts as closely as possible using both the paper and, as needed, consulting with an author of Ward et al.

## 3.2. Raw Data

The next step was identifying the data which are used for training and the search space. The provided training set, version 1.0 of Open Quantum Materials Database (OQMD) [20], [21], contains ~300,000 crystalline compounds and their properties (energy, bandgap, etc) computed via density functional theory. The original test dataset, composed of approximately 4,500 yet-undiscovered ternary compounds predicted to be stable by Meredig et al. [22], was neither openly accessible, nor made available via Ward et al. However, since the only predictions for the five most promising compound (table 1) were presented in the original work, they will be our test set.

## 3.3. Replication with Available Scripts

Following the resolution of computational dependency issues, we created models based on the *model building* script provided in the SI. Our replicated *extensibility analysis* is separated into two parts, the first script calculates and exports the descriptors for the provided training set and our test set. All the descriptors are created from the provided raw data, theoretically cloning the descriptors used in the original work. However, without a record to compare with directly or mention in the manuscript of the number of entries after data cleaning, validation of the replicated descriptors is not possible. The second script simply to retrains a model from the *model building* script using the training dataset and subsequently generate predictions for the test dataset. Despite our best efforts, the bandgaps predicted by our model deviate significantly from the reported values (see Figure 1). One potential explanation of the deviation is that our script is correct, and the differences are due to randomness in the underlying algorithms. There is no record of the random seed in the SI, so performed a random seed stability test with ten models, each with a different random seed, ensuring to include the default random seed of the underlying machine learning library, Weka [23], to determine whether the differences are due to different random initializations.

Each of the ten models created was retrained using the descriptors generated in the first script to predict the five entries within the test dataset. Figure 1 illustrates the model predictions through a violin plot, contrasting the original predictions in red with the replicated predictions, which use the default random seed, in green and the random seed stability results in blue.

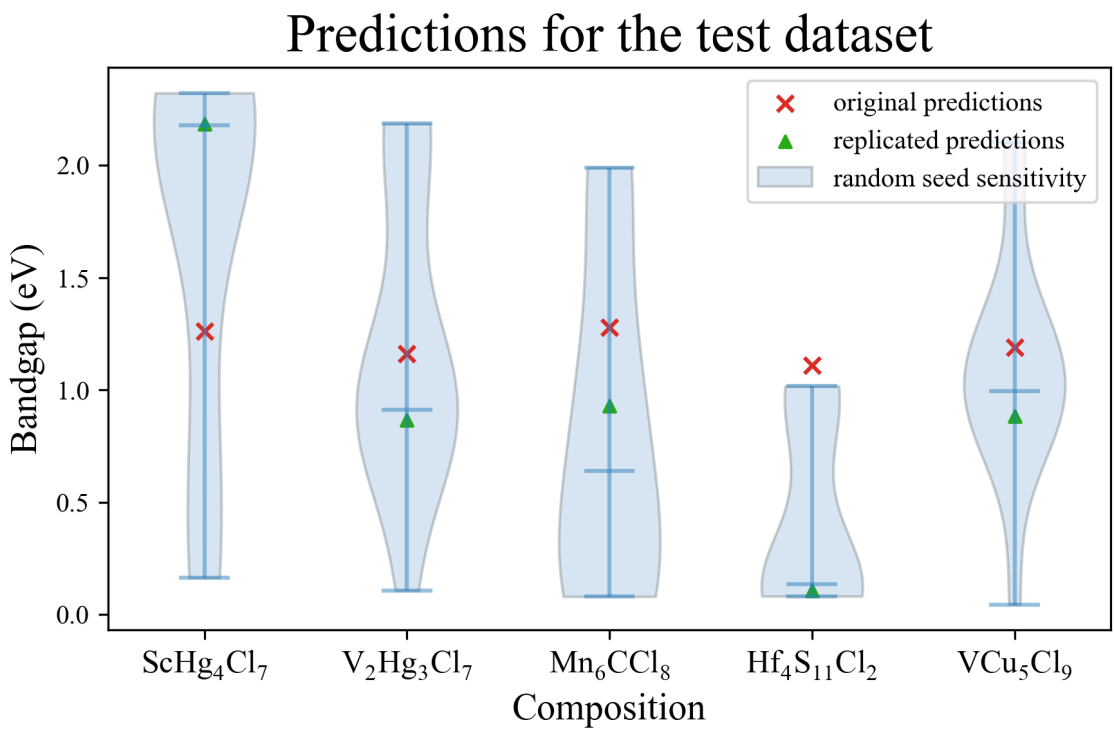

*Figure 1: The original predictions (red x's) compared to the replicated predictions (green triangles) and the predictions from the random seed sensitivity (blue violins) These result shows that across 10 different random seeds, including the default Weka seed, the original results cannot be replicated. Note: the spread in the violins is from the same model with 10 different random seeds.*

These results demonstrate that the original predictions cannot be replicated. Most of the original predictions deviate toward the tail end of the replicated predictions, with the original prediction for $Hf_4S_{11}Cl_2$ falling outside the spread of all its replicated predictions. The pronounced spread in replicated predictions emphasized sensitivity to random seed, reinforcing that in the absence of knowledge regarding the original random seed, replication becomes highly uncertain.

Our primary hypothesis is that the Magpie codebase was being continually adjusted at the time of publishing, so it is not clear whether the version in the SI is the same used to obtain the original results. The versioning issue was noted by an author as early as 2017 [24]. Without any documentation regarding the difference in Magpie versions or any of the intermediate results (descriptors, models, etc.), we cannot resolve this reproducibility challenge, as older versions are missing or have been overwritten.

## 4. Discussion

The framework proposed by Ward et al. achieved the objective of developing a generalized implementation of MI to aid materials research, while also demonstrating an effort to observe open science principles. Since the initial publication, Magpie has evolved to incorporate structural descriptors and has been integrated into Matminer, where it is now considered a standard MI baseline [16]. However, the findings presented in this study indicate a crucial lesson: reproducibility demands deliberate effort, and without it, replication becomes very difficult. As a result, the following sections detail suggestions for developers of MI tools to aid reproducibility and implicitly, the ease of use for other researchers.

### 4.1. Disseminate Dependencies

A significant hurdle to reproducibility emerges when developers presume that dependent software packages will remain compatible and readily accessible for years to come. However, the challenges posed by evolving, superseding, or abandoned dependencies are a well-recognized issues in computational reproducibility [25], [26] with many potential solutions, each accompanied by drawbacks. Docker , for instance, has gained popularity as an open platform for sharing and running tools within a loosely isolated environment to circumvent issues related to dependency management [27]. However, it requires root access privileges, posing potential security concerns for large institutions or high-performance clusters. Singularity, an alternative solution that can be used without root access, but it is limited to Linux systems [28]. Neither of these options are complete on their own, but either of them would have made our replication attempt easier. Going forward, developers must adopt a proactive stance in addressing dependency related issue, ensuring robust reproducibility when disseminating new MI tools, possibly by the containerization strategies we note above.

### 4.2. Maintain and Track Versions

In 2017, Ward published a GitHub repository [24] with an attempt to replicate the results of the original paper, noting that Magpie was in development at the same time published results were being gathered. Since there is no record of the Magpie version numbers at that time, nor a log of the machine learning operations (MLOps) for the results, it is very difficult to identify and address the root cause of our replicated prediction failing to align with the original. Employing a version control system, such as Git to track changes and the evolving state of a working directory over time aids reproducibility because changes can be easily identified and reverted [29]. A step beyond Git tracking alone would be to record all vital aspects of an experiment systematically and meticulously on a run-to-run basis.  MLFlow is a popular MLOps package which contains a tracking component to easily log numerous different aspects (start time, parameters, code version, metrics, etc.) of a project, ensuring the traceability of model lineage and fostering reproducible experiments  [30].

### 4.3. Utilize Sequential Coding Practices

Most of the example scripts Ward et al. provided were in the form of a single, extensive input script that executed many tasks and demands a large learning-curve for new users, even with the inline comments. Implementing sequential coding practices that break the study into smaller subparts is an alternative approach that increases clarity, allowing new users to appreciate how new tools work more easily. The use of coding notebooks, for instance Jupyter notebooks [31], allow developers to separate code into task-specific cells with markdown blocks in-between for comprehensive subtask descriptions. This allows a new user to identify, understand, and implement a section of interest quickly. Moreover, we propose that examples should be formed by short and concise notebooks, each representing a fundamental step, allowing for each section to be executed independently, with the outputs exported for user validation. Although implementing these practices may require more effort from developers, they are valuable to users of varying skill levels, thereby fostering quicker and better implementation.

### 4.4. Code Clarity

Reproducibility inherently requires access to all relevant code and data; the code for the extensibility analysis was not provided by Ward et al., and one of the challenges with replication is the ambiguities in the original manuscript's procedures (as described in section 3). To address this issue, we suggest incorporating clickable pointers in the manuscript that direct readers to specific files or lines of code, similar to how reference numbers are linked to citations in the reference section. Although this requires additional effort from the authors and publishers, the functionality for clickable references already exists and can greatly enhance the ability of researchers to understand the connection between the written rationale and the practical implementation.

### 5. Conclusion

In this study we attempted to reproduce task presented in "*A general-purpose machine learning framework for predicting properties of inorganic materials*" by Ward et al., where they demonstrated the application of their framework to identify materials suitable for solar cell applications. Reproduction was not possible because of incomplete code, prompting us to replicate the results while highlighting challenges encountered and their resolutions. Facilitating a straightforward way to recreate computational environment alleviates the complexities of "dependency hell," making it easier for users to reproduce results consistently. The use of version control system or dedicated tracking packages serves as a powerful mechanism for establishing a transparent and trustworthy record of the development process, instilling confidence in the reliability of published results. The division of a comprehensive script into discrete, self-contained segments, such as data cleaning, enhances user comprehension across various skill levels, enabling verification at distinct checkpoints. Lastly, incorporating clickable pointers from the manuscript to relevant files enhances clarity and reduces uncertainties inherent in concise writing. By adopting these practices, we anticipate that new tools will be more readily adopted and deployed, fostering further advancements in computational materials research.

# References

[1] National Academies of Sciences, Engineering, and Medicine (U.S.), National Academies of Sciences, Engineering, and Medicine (U.S.), National Academies of Sciences, Engineering, and Medicine (U.S.), and National Academies of Sciences, Engineering, and Medicine (U.S.), Eds., *Open science by design: realizing a vision for 21st century research*. in A consensus study report. Washington, DC: The National Academies Press, 2018.

[2] M. B. A. McDermott, S. Wang, N. Marinsek, R. Ranganath, L. Foschini, and M. Ghassemi, "Reproducibility in machine learning for health research: Still a ways to go," *Sci. Transl. Med.*, vol. 13, no. 586, p. eabb1655, Mar. 2021, doi: 10.1126/scitranslmed.abb1655.

[3] O. E. Gundersen and S. Kjensmo, "State of the Art: Reproducibility in Artificial Intelligence," *Proc. AAAI Conf. Artif. Intell.*, vol. 32, no. 1, Apr. 2018, doi: 10.1609/aaai.v32i1.11503.

[4] Committee, *Reproducibility and Replicability in Science*. Washington, D.C.: National Academies Press, 2019, p. 25303. doi: 10.17226/25303.

[5] A. Agrawal and A. Choudhary, "Perspective: Materials informatics and big data: Realization of the 'fourth paradigm' of science in materials science," *APL Mater.*, vol. 4, no. 5, p. 053208, May 2016, doi: 10.1063/1.4946894.

[6] R. K. Vasudevan *et al.*, "Materials science in the artificial intelligence age: high-throughput library generation, machine learning, and a pathway from correlations to the underpinning physics," *MRS Commun.*, vol. 9, no. 3, pp. 821–838, Sep. 2019, doi: 10.1557/mrc.2019.95.

[7] R. Ramprasad, R. Batra, G. Pilania, A. Mannodi-Kanakkithodi, and C. Kim, "Machine learning in materials informatics: recent applications and prospects," *Npj Comput. Mater.*, vol. 3, no. 1, p. 54, Dec. 2017, doi: 10.1038/s41524-017-0056-5.

[8] K. Rajan, "Materials Informatics: The Materials 'Gene' and Big Data," *Annu. Rev. Mater. Res.*, vol. 45, no. 1, pp. 153–169, Jul. 2015, doi: 10.1146/annurev-matsci-070214-021132.

[9] J. Schmidt, M. R. G. Marques, S. Botti, and M. A. L. Marques, "Recent advances and applications of machine learning in solid-state materials science," *Npj Comput. Mater.*, vol. 5, no. 1, p. 83, Aug. 2019, doi: 10.1038/s41524-019-0221-0.

[10] T. Mueller, A. G. Kusne, and R. Ramprasad, "Machine Learning in Materials Science: Recent Progress and Emerging Applications," in *Reviews in Computational Chemistry*, A. L. Parrill and K. B. Lipkowitz, Eds., Hoboken, NJ: John Wiley & Sons, Inc, 2016, pp. 186–273. doi: 10.1002/9781119148739.ch4.

[11] Y. Liu *et al.*, "Machine learning in materials genome initiative: A review," *J. Mater. Sci. Technol.*, vol. 57, pp. 113–122, Nov. 2020, doi: 10.1016/j.jmst.2020.01.067.

[12] C. Chen, Y. Zuo, W. Ye, X. Li, Z. Deng, and S. P. Ong, "A Critical Review of Machine Learning of Energy Materials," *Adv. Energy Mater.*, vol. 10, no. 8, p. 1903242, Feb. 2020, doi: 10.1002/aenm.201903242.

[13] K. Choudhary *et al.*, "Recent advances and applications of deep learning methods in materials science," *Npj Comput. Mater.*, vol. 8, no. 1, p. 59, Apr. 2022, doi: 10.1038/s41524-022-00734-6.

[14] L. Ward, A. Agrawal, A. Choudhary, and C. Wolverton, "A general-purpose machine learning framework for predicting properties of inorganic materials," *Npj Comput. Mater.*, vol. 2, no. 1, p. 16028, Aug. 2016, doi: 10.1038/npjcompumats.2016.28.

[15] K. Choudhary *et al.*, "The joint automated repository for various integrated simulations (JARVIS) for data-driven materials design," *Npj Comput. Mater.*, vol. 6, no. 1, p. 173, Nov. 2020, doi: 10.1038/s41524-020-00440-1.

[16] A. Dunn, Q. Wang, A. Ganose, D. Dopp, and A. Jain, "Benchmarking materials property prediction methods: the Matbench test set and Automatminer reference algorithm," *Npj Comput. Mater.*, vol. 6, no. 1, p. 138, Sep. 2020, doi: 10.1038/s41524-020-00406-3.

[17] L. Ward *et al.*, "Matminer: An open source toolkit for materials data mining," *Comput. Mater. Sci.*, vol. 152, pp. 60–69, Sep. 2018, doi: 10.1016/j.commatsci.2018.05.018.

[18] D. Danks and A. J. London, "Algorithmic Bias in Autonomous Systems.," in *26th International Joint Conference on Artificial Intelligence (IJCAI 2017) forthcoming*, 2017, pp. 4691--4697. [Online]. Available: https://www.cmu.edu/dietrich/philosophy/docs/london/IJCAI17-AlgorithmicBias-Distrib.pdf

[19] K. Arnold, J. Gosling, and D. Holmes, *The Java programming language*, 4th ed. Upper Saddle River, NJ: Addison-Wesley, 2006.

[20] S. Kirklin *et al.*, "The Open Quantum Materials Database (OQMD): assessing the accuracy of DFT formation energies," *Npj Comput. Mater.*, vol. 1, no. 1, p. 15010, Dec. 2015, doi: 10.1038/npjcompumats.2015.10.

[21] J. E. Saal, S. Kirklin, M. Aykol, B. Meredig, and C. Wolverton, "Materials Design and Discovery with High-Throughput Density Functional Theory: The Open Quantum Materials Database (OQMD)," *JOM*, vol. 65, no. 11, pp. 1501–1509, Nov. 2013, doi: 10.1007/s11837-013-0755-4.

[22] B. Meredig *et al.*, "Combinatorial screening for new materials in unconstrained composition space with machine learning," *Phys. Rev. B*, vol. 89, no. 9, p. 094104, Mar. 2014, doi: 10.1103/PhysRevB.89.094104.

[23] E. Frank, M. A. Hall, and I. H. Witten, "The WEKA Workbench," in *The WEKA Workbench*, 4th ed., Morgan Kaufhann, 2016.

[24] L. Ward, "identify-solar-cell-materials.ipynb." in ward-npj-2016-examples. 2017. [Online]. Available: https://github.com/WardLT/ward-npj-2016-examples/blob/master/predicting-band-gap-energies/identify-solar-cell-materials.ipynb

[25] S. L. Sawchuk and S. Khair, "Computational Reproducibility: A Practical Framework for Data Curators," *J. EScience Librariansh.*, vol. 10, no. 3, p. 1206, Aug. 2021, doi: 10.7191/jeslib.2021.1206.

[26] B. Grüning *et al.*, "Practical Computational Reproducibility in the Life Sciences," *Cell Syst.*, vol. 6, no. 6, pp. 631–635, Jun. 2018, doi: 10.1016/j.cels.2018.03.014.

[27] D. Merkel, "Docker: Lightweight Linux Containers for Consistent Development and Deployment," *Linux J*, vol. 239, no. 2, p. 2, May 2014.

[28] G. M. Kurtzer, V. Sochat, and M. W. Bauer, "Singularity: Scientific containers for mobility of compute," *PLOS ONE*, vol. 12, no. 5, p. e0177459, May 2017, doi: 10.1371/journal.pone.0177459.

[29] S. Chacon, *Pro Git*, Second edition. in The expert's voice in software development. New York, NY: Apress, 2014.

[30] A. Chen *et al.*, "Developments in MLflow: A System to Accelerate the Machine Learning Lifecycle," in *Proceedings of the Fourth International Workshop on Data Management for End-to-End Machine Learning*, Portland OR USA: ACM, Jun. 2020, pp. 1–4. doi: 10.1145/3399579.3399867.

[31] T. Kluyver *et al.*, "Jupyter Notebooks – a publishing format for reproducible computational workflows," *Elpub*, vol. 2016, pp. 87--90, 2016, doi: 10.3233/978-1-61499-649-1-87.

**Code and Data Availability**

All data and code used in this work is available on GitHub at (to be inserted upon acceptance of the paper).

**Acknowledgements**

We acknowledgements funding from Natural Sciences and Engineering Research Council of Canada, grant #RGPIN-2023-04843.

**Author Contribution**

DP conducted the experiments, analysed the results, and drafted the manuscript. LW aided in recreating scripts. All authors reviewed and edited the manuscript and contributed to manuscript preparation.